\documentstyle[12pt]{article}

\newcommand{\NP}{{\em Nucl.\ Phys.\ }}

\catcode`\@=11
\def\@citex[#1]#2{%
\if@filesw \immediate \write \@auxout {\string \citation {#2}}\fi
\@tempcntb\m@ne \let\@h@ld\relax \def\@citea{}%
\@cite{%
  \@for \@citeb:=#2\do {%
    \@ifundefined {b@\@citeb}%
      {\@h@ld\@citea\@tempcntb\m@ne{\bf ?}%
      \@warning {Citation `\@citeb ' on page \thepage \space undefined}}%
      {\@tempcnta\@tempcntb \advance\@tempcnta\@ne%
      \@tempcntb\number\csname b@\@citeb \endcsname \relax%
      \ifnum\@tempcnta=\@tempcntb %   Number follows previous--hold on to it
        \ifx\@h@ld\relax%
          \edef \@h@ld{\@citea\csname b@\@citeb\endcsname}%
        \else%
          \edef\@h@ld{\ifmmode{-}\else--\fi\csname b@\@citeb\endcsname}%
        \fi%
      \else%   %  non-successor--dump what's held and do this one
        \@h@ld\@citea\csname b@\@citeb \endcsname%
        \let\@h@ld\relax%
      \fi}%
    \def\@citea{,\penalty\@highpenalty\,}%
  }\@h@ld
}{#1}}

\def\@citeb#1#2{{[#1]\if@tempswa , #2\fi}}
\def\@citeu#1#2{{$^{#1}$\if@tempswa , #2\fi }}
\def\@citep#1#2{{#1\if@tempswa , #2\fi}}

\def\bcites{         % cite with []'s
        \catcode`\@=11
        \let\@cite=\@citeb
        \catcode`\@=12
}

\def\upcites{         % cite with exponents
        \catcode`\@=11
        \let\@cite=\@citeu
        \catcode`\@=12
}

\def\plaincites{      % cite without brackets
        \catcode`\@=11
        \let\@cite=\@citep
        \catcode`\@=12
}

\newcount\hour
\newcount\minute
\newtoks\amorpm
\hour=\time\divide\hour by 60
\minute=\time{\multiply\hour by 60 \global\advance\minute by-\hour}
\edef\standardtime{{\ifnum\hour<12 \global\amorpm={am}%
        \else\global\amorpm={pm}\advance\hour by-12 \fi
        \ifnum\hour=0 \hour=12 \fi
        \number\hour:\ifnum\minute<10 0\fi\number\minute\the\amorpm}}
\edef\militarytime{\number\hour:\ifnum\minute<10 0\fi\number\minute}

\def\draftlabel#1{{\@bsphack\if@filesw {\let\thepage\relax
   \xdef\@gtempa{\write\@auxout{\string
      \newlabel{#1}{{\@currentlabel}{\thepage}}}}}\@gtempa
   \if@nobreak \ifvmode\nobreak\fi\fi\fi\@esphack}
        \gdef\@eqnlabel{#1}}
\def\@eqnlabel{}
\def\@vacuum{}
\def\marginnote#1{}
\def\draftmarginnote#1{\marginpar{\raggedright\scriptsize\tt#1}}
\def\draft{
        \pagestyle{plain}
        \overfullrule=2pt
        \oddsidemargin -.5truein
        \def\@oddhead{\sl \phantom{\today\quad\militarytime} \hfil
        \smash{\Large\sl DRAFT} \hfil \today\quad\militarytime}
        \let\@evenhead\@oddhead
        \let\label=\draftlabel
        \let\marginnote=\draftmarginnote
        \def\ps@empty{\let\@mkboth\@gobbletwo
        \def\@oddfoot{\hfil \smash{\Large\sl DRAFT} \hfil}
        \let\@evenfoot\@oddhead}
        \def\@eqnnum{(\theequation)\rlap{\kern\marginparsep\tt\@eqnlabel}%
        \global\let\@eqnlabel\@vacuum}  }

\def\blackfonts{
        \font\blackboard=msbm10 scaled\magstep1
        \font\blackboards=msbm8
        \font\blackboardss=msbm6
}

\def\nblack{            % For people without blackboard fonts
        \def\ZZ{{Z \n{10} Z}}
        \def\NN{{N \n{14} N}}
        \def\CC{{C \n{11} C}}
        \def\RR{{R \n{11} R}}
        \def\QQ{{Q \n{12} Q}}
        \def\PP{{P \n{11} P}}
}

\def\prep{         % twocolumn.sty  Changed by Marek and Neil
        \catcode`\@=11
        \input art10.sty
        \catcode`\@=12
        
        \let\small\null
        \def\blackfonts{
                \font\blackboard=msbm10
                \font\blackboards=msbm7
                \font\blackboardss=msbm5
        }
        \let\sl\it
        \twocolumn
        \sloppy
        \voffset=-2.54truecm
        \hoffset=-2.54truecm
        \flushbottom
        \parindent 1em
        \leftmargini 2em
        \leftmarginv .5em
        \leftmarginvi .5em
        \marginparwidth 48pt
        \marginparsep 10pt
        \setlength{\columnsep}{2truecm}
        \setlength{\textwidth}{25.4truecm}
        \setlength{\textheight}{17truecm}
        \baselineskip=16pt
        \oddsidemargin .18truein
        \evensidemargin .17truein
}

\def\eqalign#1{\null\,\vcenter{\openup\jot\m@th
  \ialign{\strut\hfil$\displaystyle{##}$&$\displaystyle{{}##}$\hfil
      \crcr#1\crcr}}\,}
\def\eqalignno#1{\displ@y \tabskip\centering
  \halign to\displaywidth{\hfil$\@lign\displaystyle{##}$\tabskip\z@skip
    &$\@lign\displaystyle{{}##}$\hfil\tabskip\centering
    &\llap{$\@lign##$}\tabskip\z@skip\crcr
    #1\crcr}}

\def\section{\@startsection {section}{1}{\z@}{3.ex plus 1ex minus
 .2ex}{2.ex plus .2ex}{\large\bf}}
\def\subsection{\@startsection{subsection}{2}{\z@}{2.75ex plus 1ex minus
 .2ex}{1.5ex plus .2ex}{\bf}}
\def\subsect#1{\par\penalty1000{\noindent \bf #1}\par\penalty500}
\def\appendix{{\newpage\section*{Appendices}}\let\appendix\section%
        {\setcounter{section}{0}
        \gdef\thesection{\Alph{section}}}\section}

\def\abstract{\if@twocolumn
\section*{Abstract}
\else %\small
\begin{center}
{\bf Abstract\vspace{-.5em}\vspace{0pt}}
\end{center}
\quotation
\fi}

\catcode`\@=12

\def\noj#1,#2,{{\bf #1} (19#2)\ }
\def\jou#1,#2,#3,{{\sl #1\/ }{\bf #2} (19#3)\ }
\def\ann#1,#2,{{\sl Ann.\ Physics\/ }{\bf #1} (19#2)\ }
\def\cmp#1,#2,{{\sl Comm.\ Math.\ Phys.\/ }{\bf #1} (19#2)\ }
\def\cq#1,#2,{{\sl Class.\ Quantum Grav.\/ }{\bf #1} (19#2)\ }
\def\cqg#1,#2,{{\sl Class.\ Quantum Grav.\/ }{\bf #1} (19#2)\ }
\def\ijmp#1,#2,{{\sl Int.\ J.\ Mod.\ Phys.\/ }{\bf A#1} (19#2)\ }
\def\jmp#1,#2,{{\sl J.\ Math.\ Phys.\/ }{\bf #1} (19#2)\ }
\def\grg#1,#2,{{\sl Gen.\ Rel.\ Grav.\/ }{\bf #1} (19#2)\ }
\def\mpl#1,#2,{{\sl Mod.\ Phys.\ Lett.\/ }{\bf A#1} (19#2)\ }
\def\nc#1,#2,{{\sl Nuovo Cim.\/ }{\bf #1} (19#2)\ }
\def\np#1,#2,{{\sl Nucl.\ Phys.\/ }{\bf B#1} (19#2)\ }
\def\pl#1,#2,{{\sl Phys.\ Lett.\/ }{\bf #1B} (19#2)\ }
\def\pla#1,#2,{{\sl Phys.\ Lett.\/ }{\bf #1A} (19#2)\ }
\def\pr#1,#2,{{\sl Phys.\ Rev.\/ }{\bf #1} (19#2)\ }
\def\prd#1,#2,{{\sl Phys.\ Rev.\/ }{\bf D#1} (19#2)\ }
\def\prl#1,#2,{{\sl Phys.\ Rev.\ Lett.\/ }{\bf #1} (19#2)\ }
\def\prp#1,#2,{{\sl Phys.\ Rept.\/ }{\bf #1C} (19#2)\ }
\def\ptp#1,#2,{{\sl Prog.\ Theor.\ Phys.\/ }{\bf #1} (19#2)\ }
\def\ptpsup#1,#2,{{\sl Prog.\ Theor.\ Phys.\/ Suppl.\/ }{\bf #1} (19#2)\ }
\def\rmp#1,#2,{{\sl Rev.\ Mod.\ Phys.\/ }{\bf #1} (19#2)\ }
\def\yadfiz#1,#2,#3[#4,#5]{{\sl Yad.\ Fiz.\/ }{\bf #1} (19#2) #3%
\ [{\sl Sov.\ J.\ Nucl.\ Phys.\/ }{\bf #4} (19#2) #5]}
\def\zh#1,#2,#3[#4,#5]{{\sl Zh.\ Exp.\ Theor.\ Fiz.\/ }{\bf #1} (19#2) #3%
\ [{\sl Sov.\ Phys.\ JETP\/ }{\bf #4} (19#2) #5]}

\def\beq{\begin{equation}}
\def\eeq{\end{equation}}
\def\beqar{\begin{eqnarray}}
\def\eeqar{\end{eqnarray}}

\def\nfrac#1#2{{\displaystyle{\vphantom1\smash{\lower.5ex\hbox{\small$#1$}}%
        \over\vphantom1\smash{\raise.25ex\hbox{\small$#2$}}}}}

\def\n#1{\mskip-#1mu}

\def\lae{\mathrel{\mathop{\smash{\lower .5 ex \hbox{$\stackrel<\sim$}}}}}
\def\lae{\mathrel{\mathop{\smash{\lower .5 ex \hbox{$\stackrel>\sim$}}}}}

\def\ket#1{\left| #1 \right\rangle}

\def\l:{\mathopen{:}\,}
\def\r:{\,\mathclose{:}}

\def\[{\left[}          \def\]{\right]}
\def\({\left(}          \def\){\right)}
\def\<{\left<}          \def\>{\right>}
\catcode`\@=11
\def\theequation{\arabic{equation}}
\catcode`\@=12

\nblack
\bcites

\nblack

\catcode`\@=11
\def\theequation{\thesection.\arabic{equation}}
\@addtoreset{equation}{section}
\@addtoreset{footnote}{section}
\@addtoreset{footnote}{subsection}
\catcode`\@=12

\newcommand{\beqn}{\begin{equation}}
\newcommand{\eeqn}{\end{equation}}
\newcommand{\beqnarray}{\begin{eqnarray}}
\newcommand{\eeqnarray}{\end{eqnarray}}
\begin{document}
\begin{titlepage}

\begin{center}
%\today
June 1996
\hfill       IASSNS-HEP-96/68 \\
\hfill       PUPT-1630 \\
\hfill                  hep-th/9606136

\vskip 1 cm
{\large \bf On Tensionless Strings in $3+1$ Dimensions\\}
\vskip 0.1 cm
\vskip 0.5 cm
{Amihay Hanany%\footnote{hanany@sns.ias.edu}
}
\vskip 0.2cm
{\sl
hanany@sns.ias.edu \\
School of Natural Sciences \\
Institute for Advanced Study \\
Olden Lane, Princeton, NJ 08540, USA
}
\vskip 0.5 cm
{Igor R. Klebanov
}
\vskip 0.2cm
{\sl
klebanov@puhep1.Princeton.edu \\
Joseph Henry Laboratories \\
Princeton University \\
Princeton, NJ 08544, USA
}

\end{center}

\vskip 0.5 cm
\begin{abstract}
We argue for the existence of phase transitions in $3+1$ dimensions
associated with the appearance of tensionless strings.
The massless spectrum of this theory does not contain a graviton:
it consists of one $N=2$
vector multiplet and one linear multiplet, in agreement with the
light-cone analysis of the Green-Schwarz string in $3+1$ dimensions.
In M-theory the string decoupled from gravity arises when two 5-branes 
intersect over a three-dimensional hyperplane. The two 5-branes may be
connected by a 2-brane, whose boundary becomes a tensionless string
with $N=2$ supersymmetry in $3+1$ dimensions.
Non-critical strings on the intersection may also
come from dynamical
5-branes intersecting the two 5-branes over a string and wrapped
over a four-torus.
The near-extremal entropy of the intersecting 5-branes
is explained by the non-critical strings originating from the
wrapped 5-branes.
\end{abstract}

\end{titlepage}

\section{Introduction}

Tensionless strings have received much attention recently.
They appear very naturally in six dimensions as non-trivial infrared
fixed points.
There are various infrared data which distinguish between different
theories.

The six-dimensional tensionless string theory may carry
either the $(0, 2)$ (minimal) or the
$(0, 4)$ (which we also call $N=2$)
supersymmetry. The tensionless strings with $(0, 4)$
supersymmetry arise 
in compactification of type IIB on $K3$ 
\cite{witten} when the $K3$ gets an ADE singularity.
The self-dual D3-branes of the
type IIB theory wrap around the two cycles with small
area producing the tensionless strings. A dual description of this
is the M-theory on $T^5/Z^2$ \cite{mukhi,edfive}, where the
tensionless strings arise when two or more parallel five branes coincide
\cite{strom}. The connection between the two descriptions was found in
\cite{edfive}.

Six-dimensional tensionless strings with
$(0,2)$ supersymmetry are found
when a heterotic $E_8\times E_8$ instanton shrinks 
to zero size \cite{gh}.
The M-theory on $S^1/Z_2$ description of this
effect is provided by a 5-brane attached to a
boundary of spacetime (a 9-brane).
More general examples of tensionless strings were constructed in
\cite{sw}. For example, they arise at
the strong coupling singularity of the heterotic string on K3
\cite{sw,dlp}.

In F-theory compactifications one finds tensioneless strings
when a two cycle shrinks to zero size \cite{mv,wittenFM}. 
Its self-intersection number
serves as a quantum number which distinguishes between different types
of tensionless strings. In compactification of F-theory on Hirzebruch
surfaces $F_n$ the intersection number is $-n$.
This gives $(0,2)$
supersymmetry for $n\not=2$, and $(0,4)$ supersymmetry for $n=2$. 
An interesting approach to solving the dynamics of tensionless
strings using surface equations was proposed in \cite{ori}.

In this paper we study tensionless strings in $3+1$ dimensions.
Their existence in the context of
F-theory was pointed out in \cite{wittenFM}.
If type IIB theory is compactified on a Calabi-Yau manifold,
a tensionless string appears when a two-cycle shrinks to zero area
with a 3-brane wrapped around it. From the $3+1$ dimensional point of
view this tensionless string has $N=2$ supersymmetry.
In this paper we consider an M-theory description of such a theory,
which is provided by two 5-branes intersecting over a 3-dimensional
hyperplane. We show that the massless spectrum 
is in agreement
with the light-cone analysis of the $3+1$ dimensional Green-Schwarz
string (see section 3). 
Compactification to $2+1$ dimensions leads to a 
$N=4$ abelian gauge theory which
we study using D-brane methods in setion 4.
Going back to $3+1$ dimensions, we find an $N=2$ linear multiplet
interacting with strings.
In section 5 we show that the near-extremal entropy of the 
intersecting 5-branes may be explained by a thermal ensemble of
strings on the $3+1$ dimensional intersection.

\section{From M-theory to Tensionless Strings }

M-theory is the hypothetical unification of several types of
10-dimensional strings \cite{ht,wit}. Its low-energy effective description is
the 11-dimensional supergravity, and some valuable information can
be extracted from its classical solutions. The basic dynamical
objects of the M-theory are the 2-brane, which is electrically
charged under the 3-form gauge field, and its magnetic dual, the 5-brane.
The dynamics governing the M-brane interactions is by no means 
well-understood. Some of its features may be inferred, however, from
compactification to string theory, where the R-R charged $p$-branes
have a remarkably simple description in terms of the Dirichlet (D-)
branes \cite{dai,pol,pcj}.

The D-branes are objects on which the fundamental strings 
are allowed to end. There is evidence that a similar phenomenon
takes place in M-theory: the fundamental 2-branes are allowed to have
boundaries on the solitonic 5-branes \cite{town,strom}. 
Thus, the 5-brane
is the D-object of M-theory. The boundary of a 2-brane is a string, and
the resulting boundary dynamics appears to reduce to a kind of string
theory defined on the $5+1$ dimensional world volume.
This picture has a number of interesting implications.
Consider, for instance, two parallel 5-branes with a 2-brane stretched 
between them \cite{strom}. 
The two boundaries of the 2-brane give rise to 
two strings, lying within the first and second 5-branes respectively.
The tension of these strings may be made arbitrarily small as the
5-branes are brought close together. In particular, it can be made much
smaller than the Planck scale, which implies that the effective
$5+1$ dimensional string theory is decoupled from gravity \cite{witten}.
While it is not clear how to describe such a string theory in world sheet
terms, it has been suggested that its spectrum is given by
the Green-Schwarz approach \cite{js} ($5+1$ is one of the dimensions where
the Green-Schwarz string is classically consistent).
In the limit of coincident 5-branes, we seem to find a theory of 
tensionless strings.
These strings carry $(0,4)$ supersymmetry in $5+1$ dimensions.
Strings with $(0,2)$ supersymmetry were  
explored from several different points of 
view in refs. \cite{gh,sw,dlp}.

In this note we propose that an M-theory phenomenon, similar to the
one outlined above, leads to appearance of tensionless strings in
$3+1$ dimensions. Here the relevant M-theory configuration 
involves two 5-branes intersecting over a 3-dimensional hyperplane.
This is the coincident limit of the following 11-dimensional
supergravity solution, constructed in refs. \cite{pt,at,gkt},
\begin{eqnarray}
ds^2 &=& (F_1 F_2)^{-2/3} [ F_1 F_2 (-dt^2+ dy_1^2+ dy_2^2 
+ dy_3^2)\nonumber \\
&+& F_1 (dy_4^2+ dy_5^2) + F_2 (dy_6^2+ dy_7^2)+ \sum_{s=1}^3 (dx_s)^2 ]
\nonumber \\
F_1^{-1} &=& 1+ {Q_1\over |\vec x-\vec X_1|}\ ,
\qquad
F_2^{-1} = 1+ {Q_2\over |\vec x-\vec X_2|}\ ,\nonumber \\
{\cal F}& =& 3(* dF_2^{-1}\wedge dy_4\wedge dy_5+
* dF_1^{-1}\wedge dy_6\wedge dy_7)
\label{sol}\end{eqnarray}
where $\vec X_1$ and $\vec X_2$ are the transverse positions of the
5-branes, and $\cal F$ is the 4-form field strength.
This solution describes a number of 5-branes (measured by $Q_1$)
positioned in the $(12345)$ hyperplane, and a number
of 5-branes (measured by $Q_2$) positioned in the 
$(12367)$ hyperplane.\footnote{For the most part
we will chose $Q_1=Q_2$ to correspond to a single 5-brane in each 
position.}
This classical solution preserves $1/4$ of the original $32$ supersymmetries
of the 11-dimensional supergravity \cite{pt,at,gkt}.

Now consider a 2-brane stretched between the two 5-branes.
Strings with the smallest possible tension result from the
motion of the boundaries in the $(123)$ hyperplane.
These configurations are also special because they are supersymmetric.
To maintain supersymmetry a straight
2-brane must intersect both 5-branes over a string, and this
is only possible by positioning the 2-brane in a $(\alpha i)$
plane (the possible values of $\alpha$ are $1, 2, 3$; 
the possible values of $i$ are $8, 9, 10$). The resulting 
configuration preserves $1/8$ of the original $32$ supercharges;
i.e., the presence of a long straight string breaks 
$N=2$ supersymmetry down to $N=1$, from the $3+1$ dimensional point
of view. This implies that the straight string is a BPS saturated
state whose tension is protected by supersymmetry against
quantum corrections.

The string tension 
is proportional to the transverse distance, $|\vec X_1- \vec X_2|$.
As this distance is made much smaller than the Planck length, we expect
to find a $3+1$ dimensional string theory decoupled from gravity.
Since this theory has $8$ conserved supercharges, we find
$N=2$ supersymmetry in $3+1$ dimensions.\footnote{
Note, for comparison, that the world volume
theory of parallel 5-branes has $16$ conserved supercharges,
which corresponds to $N=2$ in $5+1$ dimensions. Upon toroidal
compactification to $3+1$, we find a theory with $N=4$
supersymmetry.} Remarkably, this is precisely
the supersymmetry of the classically consistent Green-Schwarz 
superstring. In the next section we show that
the Green-Schwarz construction indeed provides valuable information
about the massless spectrum of this theory.

\section{Massless Modes of the Green-Schwarz string in $3+1$ dimensions.}

In this section we carry out a naive light-cone quantization
of the type II Green-Schwarz string. This theory is classically consistent
in $D=3, 4, 6, 10$, and has $N=2$ supersymmetry in each of these
cases. It is possible, therefore, that some features of the
$D=4$ model are relevant to the tensionless strings arising from
the intersecting 5-branes of M-theory. In this section we examine the
massless spectrum of the $D=4$ model and find that, as expected,
it contains no gravitons.

In the light-cone approach, the massless spectrum of the
Green-Schwarz string is constructed out of the left- and right-moving
fermion zero-modes, $S_0$ and $\tilde S_0$. Each of these fields
represents a massless Majorana fermion in $3+1$ dimensions.
The transformation properties under the transverse rotation group,
$SO(2)$, are labeled by the helicity. Each massless fermion has
helicity $\pm 1/2$, so that the massless spectrum is constructed
with the following 4 operators, $S_0^{\pm 1/2}$ and $\tilde S_0^{\pm 1/2}$.

As an exercise, let us first determine the spectrum of the type I
string, which is constructed out of $S_0^{\pm 1/2}$ only.
Clearly, there are only 4 states,
\begin{equation}
\ket{0}\ , \quad S_0^{\pm 1/2} \ket{0}\ , \quad S_0^{1/2} S_0^{-1/2}\ket{0}
\ .
\end{equation}
These states combine into two massless scalars and a Majorana fermion,
which, as expected, form a $N=1$ hypermultiplet.

In proceeding to the closed string case, we can again directly
enumerate the states,
\begin{equation}
\ket{0}\ , \quad S_0^{\pm 1/2} \ket{0}\ , \quad \tilde S_0^{\pm 1/2} \ket{0}
\ , \quad {\rm etc.}
\ .
\end{equation}
We find that there are not enough degrees of freedom to form 
states of helicity $\pm 2$. As expected, there are no gravitons in
the spectrum! Instead, we find states of helicity ranging from
$-1$ to $+1$. 
Altogether, we have a vector field, 6 scalars,
and 4 Majorana fermions. 
These states combine into one $N=2$ $U(1)$ vector multiplet
and one hypermultiplet.\footnote{A $N=2$ hypermultiplet
is isomorphic to a linear multiplet consisting of an antisymmetric
tensor, $B_{mn}$, and 3 scalars.} 
It is interesting, that this is also
the field content of a single $N=4$ $U(1)$ multiplet. As explained above, 
the interactions will not respect the $N=4$ supersymmetry, however.
In the next section we will show that the massless spectrum found
from the Green-Schwarz construction is in agreement with the counting
of massless modes about the intersecting 5-branes in M-theory.

One should be concerned about the lack of consistentcy of the
$D=4$ Green-Schwarz string at the quantum level. A similar lack of 
consistency of the $D=6$ theory was addressed in ref. \cite{js}.
There it was suggested
that extra world sheet degrees of freedom, half-integer
moded oscillators, should be added to restore
Lorentz invariance. This gives a consistent free string containing
the twisted sector of the orbifold $T^4/Z_2$. The absence of the
untwisted sector makes it doubtful, however, that the interacting theory
is consistent.

In principle, we could follow a similar procedure to obtain a
Lorentz-invariant free string in $D=4$. However, our primary interest here 
is in the massless spectrum, and the half-integer moded oscillators
do not affect it. The naive light-cone procedure is sufficient
to determine the massless spectrum which, as we will show,
passes a number of consistency checks.

\section{Tensionless Strings in $3+1$ Dimensions}

In this section we discuss the low-energy field theory description of the
$5\bot 5$ configuration in M-theory. We argue that there exists
a zero-energy bound state of the 5-branes, excitations above which
are tensionless strings.

\subsect{Compactification to the intersecting D4-branes.}

First we appeal to a
somewhat indirect method for
obtaining information about the $5\bot 5$ configuration
of M-theory: we compactify one of
the coordinates of the intersection three-brane
(say, $y_1$) on a circle. 
The double dimensional reduction of the M-theory 5-brane is
known to produce a D4-brane of the type IIA theory.
Therefore, upon compactification
we find two D4-branes of type IIA theory
positioned in the $(2345)$ and $(2367)$ hyperplanes.
This is useful because a lot is known about the low-energy
description of intersecting D-branes \cite{bsv,ibrane}.
More specifically, Sen \cite{sen} has found a solution of the
$1+1$ dimensional gauge theory describing two D3-branes intersecting
over a string. Since this is related by T-duality to two
D4-branes intersecting over a 2-brane, we will make some use of the
results in \cite{sen}.

When the branes do not quite intersect, it is still convenient
to think of a pair of two-branes positioned in the $(23)$ plane 
(as the transverse distance vanishes, these two 2-branes merge
into a single intersection 2-brane).
There is a natural $Z_2$ symmetry which exchanges the two 2-branes.
The $2+1$ dimensional theory which describes their world volume
dynamics has
$8$ conserved supercharges; therefore, this is an $N=4$
supersymmetric theory. 
When the 2-branes are separated along the
transverse ($8$, $9$ and $10$) directions, their effective
field theory description is given by the $N=4$ $U(1)\times U(1)$
gauge theory with two neutral hypermultiplets \cite{bsv,ibrane}.
A geometrical interpretation of these fields may be described as
follows. The scalars of the first $U(1)$ multiplet describe the
position of the first D4-brane in the 
transverse ($8$, $9$ and $10$) directions, while 
two of the scalars
in the first neutral hypermultiplet describe its position
in the $6$ and $7$ directions.
Similarly, the second $U(1)$ multiplet describes the transverse
position
of the second D4-brane, while the second neutral hypermultiplet 
contains its $4$ and $5$ coordinates.
It is quite clear that the neutral 
hypermultiplets do not participate in the dynamics because displacement of
the D4-branes in the $4$, $5$, $6$ and $7$ directions has a trivial
effect.

The only charged fields are in the massive 
hypermultiplet which contains two complex
scalars with $U(1)\times U(1)$
charges $(1, -1)$ and $(-1, 1)$ (they
correspond to boundaries of open strings
stretched between the two D4 branes). It is convenient to combine
the two $U(1)$'s into the vector (sum) and the axial (difference)
combinations. The vector $U(1)$ corresponds 
to the overall motion of the two branes.
Since all fields are neutral under the vector $U(1)$,
it decouples.
The axial $U(1)$ participates in the non-trivial dynamics because
the two complex scalars carry charges $\pm 2$.
As the transverse distance between the 4-branes is reduced,
the charged states become lighter. Classically,
they become massless at the point
where the 4-branes intersect.

Thus, we are dealing with an interacting $N=4$ field theory
containing a $U(1)$ vector multiplet coupled to one charged hypermultiplet.
This theory may be viewed as a dimensional reduction of the
$N=2$ supersymmetric theory in $3+1$ dimensions, which
contains a $U(1)$ multiplet and a charged hypermultiplet \cite{sen}.
Decomposing the content of the theory in terms of $N=1$ multiplets,
we have one $U(1)$ vector and three hypermultiplets, with charges
$0$, $2$ and $-2$.
In terms of the charged
$N=1$ chiral superfields, $\Lambda$ and $\bar\Lambda$, and the neutral
chiral superfield, $\Phi$, we have the superpotential
\begin{equation}
W_0=\Phi \Lambda \bar\Lambda
\label{sup}
\end{equation}
$\Phi$ is related to the transverse separation between the D4-branes.
When it vanishes in the classical theory, the charged fields 
become massless.
The three scalars of the $D=3$
vector multiplet have values in $T^3/Z_2$;
the $T^3$ arises because the directions $8$, $9$ and $10$ are
compactified on a torus, while the $Z_2$ factor is due to
the symmetry interchanging the two D4-branes.
The $N=4$ axial $U(1)$ multiplet also contains a vector field which,
in $D=3$, is dual to a compact scalar. 
This scalar also changes sign under the $Z_2$, which implies
that the classical moduli space on this Coulomb branch is $T^4/Z_2$.
Seiberg has argued that quantum effects turn the moduli space
into a smooth $K3$ \cite{ns}.

There is no Higgs branch which emanates from the classical singularity
at the origin. This situation is similar to what happens in the case
of parallel five branes, as discussed in \cite{sw}.
We will argue, however, that there exists a unique supersymmetric
vacuum which describes a marginal bound
state of the intersecting D4-branes. This bound state is related by
T-duality to the marginal bound state of two D3-branes intersecting over a
string which, in turn, is U-dual to winding states of the fundamental
string \cite{sen}. Since marginal bound states are difficult to study,
it is sometimes helpful to deform the problem in a way that turns it
into a true bound state with a mass gap.
The above superpotential does not allow for
such a situation so we need to modify it next.

The modification is analogous
to that used by Sen in the $1+1$ dimensional case \cite{sen}.
He introduced a constant electric field, $F_{01}$.
The necessity of screening this field in a supersymmetric vacuum drives
the theory into the Higgs phase. Physically, the electric field
corresponds to inserting some number of fundamental strings, which
form a bound state together with the intersecting D3-branes.

What is the analogue of this phenomenon in $2+1$ dimensions?
One possiblility is to introduce
a transverse magnetic field $F_{12}$. If we take its value to be
independent of all the coordinates, then the parity is broken, but not
the spatial translation invariance. 
To achieve a supersymmetric ground state,
$F_{12}$ needs to be screened. This creates an energy barrier
against taking $\Phi$ to be large, since in that case the charges
become arbitrarily massive. To determine the structure of the theory
for small $\Phi$, we follow \cite{ed,sen} and add mass terms to
the superpotential,
\begin{equation}
W=\Phi \Lambda \bar\Lambda + {1\over 2}\epsilon \Phi^2 +
\eta \Lambda \bar\Lambda
\end{equation}
The non-trivial critical point of $W$, corresponding to 
a supersymmetric vacuum, is given by
\begin{equation}
\Phi =-\eta\ , \qquad \Lambda= \bar\Lambda= \sqrt{\epsilon\eta}
\end{equation}
This is a vacuum, where the magnetic field is screened.
This phenomenon is nothing but the Meissner effect found in
superconductors. In this phase we find that the magnetic flux
is confined to the Abrikosov-Nielsen-Olesen strings.
As we comment in the next section, these are likely to be
the strings we are after.

There seems to exist another deformation of the theory that turns
the marginal bound state into a state with a mass gap.
We may add the abelian Chern-Simons term for the gauge field 
which will cause the vector multiplet to become
massive (its coefficient is the
mass of the $U(1)$ multiplet).

\subsect{The $3+1$-dimensionsional theory on the $5\bot 5$ intersection}

{}From the results above, we have some information about what happens to
the $3+1$ dimensional intersection theory upon compactification
on a circle to $2+1$ dimensions.
Now we attempt to ``undo'' the compactification and
learn more about the
exotic theory of tensionless
strings. Indeed, it is not hard to see how
the degrees of freedom in the $2+1$ dimensional theory
combine into the $N=2$ multiplets of the $3+1$ dimensional
theory.
When the 5-branes are separated, we have $N=2$ supersymmetric
$U(1)\times U(1)$ theory with two neutral 
linear multiplets.  
Note that each of the branes contributes a $N=2$ vector multiplet
and a linear multiplet, precisely in accord with the spectrum of
the $3+1$ dimensional Green-Schwarz string discussed in section 3.
Now it is natural to interpret the scalars
from the vector multiplets as corresponding to motion in the
$4$, $5$, $6$ and $7$ directions (upon dimensional reduction,
the vector multiplets reduce to the duals of the neutral 
hypermultiplets in $2+1$ dimensions).
Each $N=2$ linear multiplet 
contains $B_{mn}$ and 3 scalars.
The antisymmetric tensor, which is dual to a compact scalar (axion),
originates from
the dimensional reduction of a $D=6$ 
antiselfdual two-form field, $B_{mn}^-$. 
The 3 scalars in the first (second) linear multiplet
describe the transverse ($8$, $9$ and $10$) coordinates of the
first (second) 5-brane.
Combining the two linear multiplets into a sum and a difference,
we see that the sum corresponds to the overall transverse motion.
Only the difference multiplet participates in the dynamics.

The crucial question is: what is the $3+1$ dimensional realization
of the extra charged multiplets, which become massless
in $2+1$ dimensions in the limit when the two branes coincide.
Since in $2+1$ dimensions we found one $N=4$ hypermultiplet
containing charges $(-1, 1)$ and $(1, -1)$ under $U(1)\times U(1)$,
it is clear that in $3+1$ dimensions we have
a multiplet containing axionic
charges $(-1, 1)$ and $(1, -1)$ under the two linear multiplets.
Classically, these charges are 
carried by long straight strings.
As the strings become tensionless,
their winding modes turn into the massless charges of 
the $2+1$ dimensional compactified theory.
There are two types of tensionless strings
which
come from the two possible 
orientations of the 2-brane stretched between the 5-branes.
There is also a $Z_2$ action which exchanges the two fivebranes.

The classical moduli space of the difference linear multiplet
is parameterized by four real compact scalars. 
Three of the scalars denote the $(8,9,10)$
relative motion and change sign under the $Z_2$. 
The other compact scalar is dual to the $B_{mn}$ field and also
changes sign under the $Z_2$. So, the classical moduli space is
$T^4/Z_2$. Classically, we find tensionless strings at the
singularity of this moduli space. 
As for the $2+1$ dimensional
system analyzed above,
we expect no Higgs branch to emanate from the singularity
point. We believe, however, that there exists a
supersymmetric vacuum corresponding to a marginal bound state of
the intersecting 5-branes.

In the previous section we argued
that the problem may be deformed in such a way that the marginal bound
state turns into a state with a mass gap.
This bound state is described by the Higgs phase
of a
supersymmetric Landau-Ginzburg model
in $2+1$ dimensions where the charged fields have condensed.\footnote{
In $3+1$ dimensions this should look like condensation of
strings.} It is well-known that this model
contains string-like solitons, the Nielsen-Olesen vortices.
Let us recall their basic structure.
We will work with a $N=2$ $U(1)$ multiplet in $2+1$
dimensions interacting with a charged
hypermultiplet. In the Higgs phase, the charged scalars acquire
expectation values, and the massless $U(1)$ vector multiplet combines
with the hypermultiplet to form a massive vector multiplet.
The BPS saturated Nielsen-Olesen strings
arise from the interaction of
a charged scalar with the $U(1)$ field. The static energy
of a string of length $L$ is given by
\begin{equation}
E= L \int d^2 x \left [{1\over 4 e^2} F_{mn}^2 + |D_m \Lambda|^2 +
\lambda \left (|\Lambda|^2 - {F^2\over 2}\right)^2 \right ]
\end{equation}
$F=\sqrt{\epsilon\eta}$ is the expectation value of the charged
scalar, $\Lambda$.
If we chose $2\lambda= e^2$, so that both the scalar mass and the vector
mass are equal to $m=e F$, then there exists a Bogomolny bound on the
energy of the vortex solution \cite{EB}. 
If the vortex contains $n$ units of the
magnetic flux, then
\begin{equation}
E\ge F^2 L\pi |n|
\end{equation}
When the bound is saturated, the vortex solution
breaks only $1/2$ of the supersymmetries. 
The elementary vortices of charge $\pm 1$ 
are presumably related to the
strings that arise at the M-brane intersections.

In order to see which objects govern the low-energy dynamics of
the abelian Higgs model, we have to compare the value of the 
Nielsen-Olesen string tension to the scalar and vector mass, 
$m$.
{}From the above equations, we find
\begin{equation}
T = {m^2 \pi\over e^2}
\end{equation}
For small $e$ (weak coupling) the strings are heavy and
the model has a conventional field theoretic description.
This is the usual limit in which the abelian Higgs model is discussed.
Note, however, that for strong coupling the strings should become
more relevant because their tension becomes small.\footnote{We
are very grateful to J. Distler for illuminating discussions of
this point.} Here we may find a non-field theoretic behavior
dominated by the tensionless strings. It is tempting to
speculate that, as the theory flows to strong coupling,
the tensionless strings regulate the Landau pole
found in the purely field theoretic approach.

We believe that the Nielsen-Olesen strings that become light at
strong coupling provide
a good model for the tensionless strings that are found
in M-theory when the 5-branes approach each other.
Clearly, we need a better understanding of how this comes
about at the quantum level, but we find it
remarkable that the M-theory may provide a quantum
description of fluctuating superconducting vortex lines.

In conclusion we would like to mention a different field-theoretic scenario
for the transition that takes place when the 5-branes coincide.
The interacting system relevant
to the separated 5-branes consists of a linear multiplet coupled
to strings whose tension is proportional to the separation.
We may imagine that these strings are the Nielsen-Olesen strings
discussed above.
In other words, the strings originate
from a hidden $N=2$ supersymmetric $U(1)$
gauge theory coupled to a charged hypermultiplet in the Higgs phase.
The expectation value of the linear multiplet is identified with the
separation between the 5-branes.

We denote the scalars in the linear multiplet as a complex field
$\theta$ and a real field $r$. They transform in the adjoint
representation of the $SU(2)_R$ symmetry. There is a coupling of these
scalars to the D-terms of the gauge field which also transform under
the adjoint of $SU(2)_R$. By supersymmetry there is also a coupling of
the B field to the field strength of the gauge field of the form
$B_{mn}F^{mn}$. As a result,
the D-term constraints are
% the equations for the superpotential
%(\ref{sup}) are changed in addition to the D-term constraints.
\beq
\Lambda\bar\Lambda-\theta=0,\qquad |\Lambda|^2-|\bar\Lambda|^2=r.
\eeq
A solution exists for any value of $r$ and $\theta$. In particular, 
$\Lambda=\bar\Lambda=0$
if and only if $r=\theta=0$: namely, when the two five-branes
coincide. 

As the 5-branes coincide, the hidden $U(1)$ becomes ``un-Higgsed''
and it seems that on the other side of the phase transition we have
a Coulomb phase of $U(1)$ coupled to a charged hypermultiplet.
%However, the latter field theory is ill-defined because it is not
%asymptotically free. We expect that it is given meaning only in the
%context of string theory or M-theory. 
It is also possible
that instead of the Coulomb phase we have a single quantum state
describing the bound 5-branes.

\section{The Near-Extremal Entropy}

In this section we provide another piece of evidence for the
existence of a string theory on the $3+1$ dimensional
intersection of two M-theory 5-branes. This evidence, albeit
indirect, comes from studying the near-extremal entropy of
the intersecting 5-branes. The necessary non-extremal supergravity
solution was recently found by Cveti\v c and 
Tseytlin \cite{CT}, and we
will simply use their results.

The solution is characterized by 
the non-extremality parameter $\mu$, and the
charges $Q_1$ and $Q_2$ which
are proportional to the numbers of the $(12345)$ and 
$(12367)$ 5-branes respectively. For small $\mu$ 
the ADM mass and the Bekenstein-Hawking entropy are \cite{CT}
\begin{eqnarray}
M&=& b \big ( Q_1+Q_2+\mu +O(\mu^2) \big )
\nonumber \\
S_{BH} &=& c \mu \sqrt{Q_1 Q_2} + O(\mu^2)
\end{eqnarray}
Using $c/b=4\pi$, we have near extremality,
\begin{equation}
S_{BH} = 4\pi \sqrt{Q_1 Q_2} E\ ,
\end{equation}
where $E= M- M_0$. This implies that the Hawking temperature
is constant and does not depend on $E$, 
\begin{equation}
T_H= {1\over 4\pi \sqrt{Q_1 Q_2}}\ .
\end{equation}
A similar phenomenon
also occurs for a 5-brane in 10 dimensions and was interpreted
by Maldacena \cite{juan}
as due to a gas of strings at its Hagedorn temperature,
which is equal to the Hawking temperature.

Consider a gas of strings with tension
$T_{eff}=1/(2\pi \alpha'_{eff})$, and with world sheet degrees of
freedom having central charge $c_{eff}$. At high energy $E$,
the entropy is the same as for a single long string,
\begin{equation}
S= 2\pi \sqrt{c_{eff}\alpha'_{eff}\over 6} E
\end{equation}
Comparing with the Bekenstein-Hawking entropy, we interpret
the near-extremal entropy in terms of a single string with
\begin{equation} \label{constrain}
c_{eff}\alpha'_{eff}= 24 Q_1 Q_2
\end{equation}
Now we need to recall how $Q_1$ and $Q_2$ are related to 
$n_1$, the number of 5-branes in the $(12345)$ plane, and
$n_2$, the number of 5-branes in the $(12367)$ plane. 
This was explained
in \cite{KT}, with the result
\begin{equation}
Q_1 = {n_1\over 2\pi L_6 L_7} \left (\pi \kappa^2\over 2 \right )^{1/3}
\ ,\qquad\quad
Q_2 = {n_2\over 2\pi L_4 L_5} \left (\pi \kappa^2\over 2 \right )^{1/3}
\ .
\end{equation}
Here $L_i$ are the sizes 
of the compact dimensions, $(4, 5, 6, 7)$,
and $\kappa$ is the 11-dimensional gravitational constant.

We will assume that $c_{eff}$ does not depend on the parameters,
while the effective tension does. This leads to
\begin{equation}
T_{eff}\sim {L_4 L_5 L_6 L_7 \over n_1 n_2 \kappa^{4/3} }
\ .
\end{equation}
In other words, the effective tension is proportional to
the volume of the $(4, 5, 6, 7)$ dimensions. 
J. Maldacena suggested to us an interesting interpretation of
our formula for the string tension: the string is the 5-brane
wrapped around the $(4, 5, 6, 7)$ dimensions.
This interpretation passes some consistency checks.
For instance, if $n_1=n_2=1$ we expect the string to originate from a 
single 5-brane. Using the known value of the 5-brane tension \cite{KT},
\begin{equation}
T_5 =\left ( {\pi\over 2}\right )^{1/3} \kappa^{-4/3}
\ ,\end{equation}
we find that $T_{eff}=T_5 L_4 L_5 L_6 L_7$ is consistent with
(\ref{constrain}) provided we use $c_{eff}=6$.
For general $n_1$ and $n_2$, the effective string tension is
\begin{equation} \label{tension}
T_{eff}={T_5 L_4 L_5 L_6 L_7\over n_1 n_2}
\end{equation}
Thus, the tension of a wrapped 5-brane is reduced by a multiplicative factor.
A similar reduction of tension was
necessary for explaining the entropy of a D-string
moving within a number of parallel type IIB 5-branes \cite{juan}.

Thus, the non-critical strings relevant to
the $5\bot 5$ configuration with no transverse separation are M-theory 
5-branes wrapped around $T^4$. This may seem surprising, since for
transversely separated 5-branes the strings originate from stretched 
2-branes. When the 5-branes are coincident, and the transverse
coordinates $(8, 9, 10)$ are compact, we may, in fact, add a 2-brane
intersecting the 5-branes and wrapped around one of the transverse
dimensions (say, $8$). This would create a string in $3+1$ dimensions
with tension $T_2 L_8/(n_1 n_2)$.\footnote{
A similar configuration of M-theory
explains the effective string picture of \cite{juan}. Consider a 
5-brane solution with one of the transverse dimensions taken to be 
compact. We may then add a 2-brane intersecting
the 5-brane over a string and wrapped around the compact dimension.
Upon reduction to type IIA theory, this gives a fundamental string
moving within the NS-NS 5-brane. This is U-dual to the D-string moving
within the D5-brane, which was studied in \cite{juan}. }
However, for the classical solution (\ref{sol}) the transverse
coordinates are non-compact, and this is impossible.

Let us recall that there exists another
configuration of M-theory which preserves $1/8$ of the 
supersymmetries \cite{pt,at,gkt}:
we can add a 5-brane in the $(a4567)$ hyperplane, with $a=1, 2$ or $3$.
Thus, the additional 5-brane has a string in common with the
$(12345)$ and $(12367)$ hyperplanes. The low-energy excitations of
the additional 5-brane will make the string fluctuate within the
$(123)$ hyperplane, while its tension is given by (\ref{tension}). 
This provides a fairly consistent picture of the $3+1$ dimensional
non-critical string which arises for strictly intersecting 5-branes.
However, it remains to be explained precisely why this string has
$c_{eff}=6$.

If the volume of $T^4$, $L_4 L_5 L_6 L_7$, is
made very small in Planck units, the string seems to decouple
from 11-dimensional gravity.
Increasing $n_1$ and $n_2$ also serves to
reduce the effective tension of the $3+1$ dimensional string.
To summarize, in this section we interpreted the near-extremal entropy
of the $5\bot 5$ configuration in terms of strings on the
$3+1$ dimensional intersection. The parameters may be adjusted in such
a way that this string theory decouples from gravity, as anticipated
from our previous anlaysis. 

\section{Discussion}

In this paper we have argued that
tensionless strings are not unique to $5+1$ dimensions.
A concrete $3+1$-dimensional example is provided by the
$5\bot 5$ configuration of M-theory. The two 5-branes
may be connected by a 2-brane, whose boundary then
acts as a non-critical string.
Another source of non-critical strings is a 5-brane
which
intersects the two original 5-branes over a string
and is wrapped over $T^4$.
The tension of the non-critical
strings may be made arbitrarily small.
Thus, the graviton should
not be a part of the spectrum. Indeed, we find that the only
massless fields the string couples to lie in the vector and hyper
$N=2$ multiplets. This agrees with the light-cone analysis of the
massless spectrum of the Green-Schwarz superstring in $3+1$ dimensions.

We would also like to remark that tensionless strings may be found
in $4+1$ dimensions, simply by dimensionally reducing the 
$5+1$ dimensional case. This case has a direct
D-brane desription.
Indeed, consider the M-theory configuration consisting of two
parallel 5-branes connected by a 2-brane. Double dimensional reduction
of the 5-branes along a direction orthogonal to the 2-brane brings us
to the type IIA configuration of two parallel D4-branes connected by
a stretched D2-brane. This configuration is described by a SU(2)
monopole in the world volume $N=4$ gauge theory.
To see this, note that T-duality relates it to a pair of D3-branes
connected by a D1-brane. The relation of this to monopoles was explained
in \cite{ts,gg,dl}. Far away from the monopole the Higgs fields reaches
a constant value which measures the separation between the 3-branes;
at the core of the monopole, the Higgs field vanishes which means that
the 3-branes are connected by a D-string. Clearly, the monopole may also be 
regarded as a special solution of the $4+1$ dimensional gauge theory,
which is translationally invariant along one of the directions.
This solution describes two parallel D4-branes connected by a D2-brane.
Presumably, there also exists a gauge theory
configuration describing a 2-brane stretched between two
intersecting D4-branes,
but we leave this for future work.

In conclusion we would like to comment on the relation between
supersymmetric non-critical strings and gauge theories.
It has been proposed that, upon compactification on
$T_2$, the anti-selfdual string in $D=6$ reduces to
$N=4$ supersymmetric $SU(n)$ gauge theory in $D=4$ \cite{witten,js}.
The electric and magnetic charges in $D=4$ arise as strings wrapped
around the two different cycles of $T_2$.

There exists a similar relation between $N=2$ supersymmetric 
non-critical strings in
$D=4$ and $N=4$ gauge theory in $D=3$ obtained by compactification on
a circle. If we start in the M-theory with $n$ 5-branes in the
$(12345)$ plane and $m$ 5-branes in the $(12367)$ plane, then upon
dimensional reduction we arrive at $n$ D4-branes intersecting
$m$ D4-branes. The intersection is described by $D=3$ $N=4$
supersymmetric $U(1)\times SU(n)\times SU(m)$ gauge theory
coupled to three hypermultiplets \cite{bsv,sen}. Two of the hypermultiplets
are neutral under $U(1)$; the first is in the adjoint representation
of $SU(n)$, the second -- of $SU(m)$. The third hypermultiplet 
contains a $(n, \bar m)$ representation of $U(1)$ charge 2, and
a $(\bar n, m)$ representation of $U(1)$ charge $-2$.
In this paper we analyzed the purely abelian case of $n=m=1$.
The non-abelian dynamics is more complicated, and we hope to return to
it in the future.

\section*{Acknowledgements}
We thank J. Distler, O. Ganor, K. Intriligator, J. Maldacena,
A. Polyakov, C. Schmidhuber, N. Seiberg and E. Witten
for useful discussions. We are especially grateful to A. Tseytlin
for many discussions and explanations of his work.
A.H. was supported in part by NSF grant PHY-9513835.
I.R.K. was supported in part by DOE grant DE-FG02-91ER40671, the NSF
Presidential Young Investigator Award PHY-9157482, and the James S.{}
McDonnell Foundation grant No.{} 91-48.


\begin{thebibliography}{99}

\small
\parskip=0pt plus 2pt

\bibitem{witten}
E.~Witten, ``Some Comments On String Dynamics,'' hep-th/9507121.

\bibitem{mukhi}
K. Dasgupta and S. Mukhi, ``Orbifolds of M-Theory,''
hep-th/9512196.

\bibitem{edfive}
E.~Witten, ``Five-branes And $M$-Theory On An Orbifold,''
hep-th/9512219.

\bibitem{strom}
A.~Strominger, ``Open P-Branes,'' hep-th/9512059.

\bibitem{gh}
O.~Ganor and A.~Hanany, ``Small $E_8$ Instantons and Tensionless
Non-critical Strings,'' hep-th/9602120.

\bibitem{sw}
N.~Seiberg and E.~Witten, ``Comments on String Dynamics in Six
Dimensions,'' hep-th/9603003.

\bibitem{dlp}
M.~Duff, H~Lu and C.~Pope, ``Heterotic Phase Transitions and
Singularities of the Gauge Dyonic String,'' hep-th/9603037.

\bibitem{mv} D.~Morrison and C.~Vafa, ``Compactifications of
F-Theory on Calabi-Yau Threefolds, I and II,''
hep-th/9602114, 9603161.

\bibitem{wittenFM}
E.~Witten, ``Phase Transitions In M-Theory And F-Theory,''
hep-th/9603150.

\bibitem{ori}
O.~Ganor, ``Six Dimensional Tensionless Strings in the Large N
Limit,'' hep-th/9605201.

\bibitem{ht}
C.~M.~Hull and P.~K.~Townsend,
``Unity of Superstring Dualities,'' hep-th/9410167,
\NP {\bf B438} (1995)  109.

\bibitem{wit}
E.~Witten, ``String Theory Dynamics In Various Dimensions,''
hep-th/9503124, \NP {\bf B443} (1995)  85.

\bibitem{dai}
J.~Dai, R.~G.~Leigh and J.~Polchinski, ``New Connections Between String
Theories,'' \mpl4,89,2073.

\bibitem{pol}
J.~Polchinski, ``Dirichlet-Branes and Ramond-Ramond Charges,''
hep-th/9510017.

\bibitem{pcj}
For a recent review, see J.~Polchinski, S.~Chaudhuri,
and C.~V.~Johnson, ``Notes on D-Branes,''
NSF-ITP-96-003, hep-th/9602052.

\bibitem{town}
P.~K.~Townsend, ``D-branes from M-branes,'' hep-th/9512062.

\bibitem{js} J.~Schwarz, ``Self-Dual Superstring in Six Dimensions,''
hep-th/9604171.

\bibitem{pt}
G.~Papadopoulos and P.~K.~Townsend, ``Intersecting M-branes,''
hep-th/9603087.

\bibitem{at}
A.~A.~Tseytlin, ``Harmonic superpositions of M-branes,''
hep-th/9604035.

\bibitem{gkt}
J.~P.~Gauntlett, D.~A.~Kastor and J.~Traschen, 
``Overlapping Branes in M-Theory,''
hep-th/9604179.

\bibitem{bsv} M.~Bershadsky, V.~Sadov and C.~Vafa,
``D-Strings on D-Manifolds,'' hep-th/9510225.

\bibitem{ibrane}
M.~Green, J.~A.~Harvey and G.~Moore,
``I-Brane Inflow and Anomalous Couplings on D-Branes,'' hep-th/9605033.

\bibitem{sen}
A.~Sen, ``U-Duality and Intersecting D-Branes,'' hep-th/9511026.

\bibitem{ns}
N.~Seiberg, ``IR Dynamics on Branes and Space-Time Geometry,''
hep-th/9606017.

\bibitem{ed}
E.~Witten, ``Bound States Of Strings And $p$-Branes,'' hep-th/9510135.

\bibitem{EB}
E. Bogomolny, 
``Stability of Classical Solutions,''
{\it Yadernaya Fizika} {\bf 24} (1976) 861.

\bibitem{CT}
M.~Cveti\v c and A.~Tseytlin, ``Non-Extreme Black Holes
from Non-Extreme Intersecting M-branes,''
hep-th/9606033.

\bibitem{juan}
J.~Maldacena, ``Statistical Entropy of Near Extremal Five-Branes,''
hep-th/9605016.

\bibitem{KT}
I.~R.~Klebanov and A.~Tseytlin, ``Four-Dimensional Black Holes
as Intersecting M-branes,'' hep-th/9604166.

\bibitem{ts}
A.~A.~Tseytlin, ``Self-duality of Born-Infeld action and Dirichlet
3-brane of type IIB superstring theory,'' hep-th/9602064.

\bibitem{gg}
M.~Green and M.~Gutperle, ``Comments on Three-Branes,'' hep-th/9602077.

\bibitem{dl}
M.~Douglas and M.~Li,
``D-Brane Realization of N=2 Super Yang-Mills Theory in Four
Dimensions,'' hep-th/9604041.

\end{thebibliography}
\end{document}